# Thin Shell Model of a Coated Conductor with a Ferromagnetic Substrate

Leonid Prigozhin and Vladimir Sokolovsky

*Abstract*— Coated conductors with magnetic substrates are thin multilayer structures; their high aspect ratio and nonlinear material properties present significant difficulties for numerical simulation. Using the high width-to-thickness ratio of coated conductors we derive an integral formulation for a model based on an infinitely thin approximation for the superconducting layer and a quasistatic thin shell approximation for the magnetic substrate. The proposed model describes electromagnetic response of a coated conductor with a magnetic substrate and is much simpler than the existing models. A single dimensionless parameter characterizes the substrate having a finite magnetic permeability and a finite thickness. An accurate and efficient Chebyshev spectral method is derived for numerical solution. The influence of a magnetic substrate on the superconducting current and AC losses is investigated. In the limiting cases our model solution tends to the known analytical solutions.

## I. INTRODUCTION

Coated conductors are multilayer tapes containing, in particular, an approximately 1 $\mu m$ thick layer of a type-II superconducting material upon a metallic substrate. The substrate is much thicker, 30-100 $\mu m$, and can be made of a magnetic material, e.g., a Ni-W alloy. The substrate thickness is, however, small compared to its width, typically, 4-12 mm. Except for high frequencies and amplitudes of the applied field, the AC loss in a superconducting layer is usually much higher than losses in other parts of a coated conductor. This loss is influenced by the magnetic substrate, which changes the distribution of a superconducting current. Modeling losses and currents in coated conductors with magnetic substrates is currently an active area of research (see [1-8] and the references therein). Assuming an infinite magnetic permeability of the substrate, Mawatari ([1], see also [5]) solved the problem analytically for a superconducting layer either in the Meissner state or described by the Bean critical-state model. For problems with a finite magnetic permeability of a substrate and different current-voltage relations characterizing the superconductor, various two-dimensional (2d) formulations have served a basis for the finite element approximation and numerical solution [2, 4, 6, 8-12].

The h-formulation, employed in [6, 10, 13] and often realized in a commercial finite element software, can be not the optimal choice because of a very high width-to-thickness ratio of a conductor and the necessity to match its finite element discretization to that of the surrounding space. The same difficulty arises in the A-formulation [12, 14, 15]. A better approach is, probably, to use an integral formulation for which the finite element discretization is needed only for the active parts of the system, the superconducting and magnetic layers. Such a method has been recently presented in [7], where the equivalent surface current model was applied to a 2d integral formulation after its finite element discretization. Using one layer of rectangular elements in the superconductor and two layers of such elements in the substrate, these authors successfully solved a problem with the field-dependent both magnetic permeability of the substrate and critical current density of the superconductor.

We note, however, that for a thin ferromagnet shell or plate the equivalent surface current model presents the induced field as a difference of two close to each other fields; subtraction cancellation can negatively affect the accuracy of numerical solution. Furthermore, the current density in a thin superconducting strip is well described by a one-dimensional (1d) integral equation, so approximating its cross-section by 2d rectangular elements is usually unnecessary. Although the substrate is thicker, its aspect ratio is still high, which justifies using the thin shell magnetization model [16-18]. In the long strip case this model boils down to a 1d integro-differential equation.

Assuming, for simplicity, a constant magnetic susceptibility of the substrate, $\chi$, we derive in this work a system of singular 1d integro-differential equations for the sheet current density in the superconductor and the so-called surface magnetization [16] of the substrate. This system (with any nonlinear current-voltage relation for the superconductor) can be solved numerically using an efficient and highly accurate Chebyshev spectral method. Such methods make possible simple analytical treatment of the integral kernel singularities, converge quickly, and have recently been applied to other superconductivity problems in our works [19-21].

The 1d model proposed in this work contains a dimensionless parameter, $\kappa = \chi \delta / a$, where $\delta$ is the substrate thickness and $2a$ is the strip width. The magnetic susceptibility $\chi$ of a substrate is zero for non-magnetic materials; for magnetic substrates it is, usually, of the order of tens but can be much higher (thousands and even more [22]). Hence, there is a wide range of possible $\kappa$ values. For a superconductor in the Meissner state, our model solution tends to the known analytical solutions, [23, 24] or [1], as $\kappa \to 0$ or $\kappa \to \infty$,

V. Sokolovsky is with Physics Department, Ben-Gurion University of the Negev, Beer-Sheva 84105 Israel (sokolov@bgu.ac.il ).

Color versions of the figures in this article are available online at http://ieeexplore.ieee.org

respectively. For the power current-voltage relation with a large power, the model of a superconductor is close to the critical-state model. Solutions of our model for small and large $\kappa$ also in this case tend to the corresponding analytical solutions in [24] and [1], respectively.

Comparing to the existing models and numerical methods for the coated conductor problems with a finite value of $\kappa$ and an arbitrary nonlinear current-voltage relation characterizing the superconductor, the advantage of our 1d model and spectral method is their simplicity and efficiency without any compromise in accuracy.

Our numerical simulations were performed in Matlab R2020b on a PC with the Intel® Core™ i7-9700 CPU 3.00 GHz.

## II. THE MODEL

We assume the coated conductor is infinite in the $z$-axis direction and limit our consideration to the superconducting layer on a magnetic substrate: usually, other layers of the coated conductor do not significantly affect the conductor electromagnetic response. The substrate material is assumed to be an ideal ferromagnetic: non-conducting and characterized by a constant magnetic susceptibility $\chi \gg 1$. The substrate cross-section is $\{|x| < a, |y| < \delta/2\}$, while the superconducting layer is assumed infinitely thin, $\{|x| < a, y = \delta/2\}$. We also assume $\delta/a \ll 1$. Our aim is to derive the model in the limit $\delta \to 0$, $\chi \to \infty$ while their product $\chi\delta$ remains constant.

The magnetic field can be presented as a superposition,

$$\boldsymbol{h} = \boldsymbol{h}^{\mathrm{e}} + \boldsymbol{h}^{\mathrm{sc}} + \boldsymbol{h}^{\mathrm{fm}} \qquad (1)$$

where $\boldsymbol{h}^{\mathrm{e}}$ is the given external field, $\boldsymbol{h}^{\mathrm{sc}}$ is the field induced by the superconducting current, and $\boldsymbol{h}^{\mathrm{fm}}$ is the field generated by the magnetic substrate. Provided the external field and coated conductor properties do not vary along the $z$-axis, the sheet current density in the superconducting layer is directed along this axis and can be regarded as a scalar, $j = j(t, x)$. Neglecting the vanishing $(\delta \to 0)$ thickness of the substrate and using the Biot-Savart law we find that outside the superconducting layer

$$\boldsymbol{h}^{\mathrm{sc}}(t, x, y) = \frac{1}{2\pi} \begin{Bmatrix} -\int_{-a}^{a} \dfrac{j(t, x')y\mathrm{d}x'}{(x-x')^2 + y^2}, \\ \int_{-a}^{a} \dfrac{j(t, x')(x-x')\mathrm{d}x'}{(x-x')^2 + y^2} \end{Bmatrix}. \qquad (2)$$

The $y$-component of this field is continuous, while the $x$-component has a jump at $y = 0$, $|x| < a$:

$$h_x^{\mathrm{sc}}\big|_{y=0-} = +\frac{j}{2}, \quad h_x^{\mathrm{sc}}\big|_{y=0+} = -\frac{j}{2}. \qquad (3)$$

The thin substrate, placed just below the superconducting layer, is thus exposed to the field

$$\boldsymbol{h}^0(t, x) = \boldsymbol{h}^{\mathrm{e}}(t, x, 0) + \begin{Bmatrix} \dfrac{j}{2}, \\ \dfrac{1}{2\pi}\int_{-a}^{a} \dfrac{j(t, x')\mathrm{d}x'}{x - x'} \end{Bmatrix}. \qquad (4)$$

To relate this field to the field $\boldsymbol{h}^{\mathrm{fm}}$ generated by the magnetic substrate, we use the thin shell magnetization theory developed by Krasnov [16-18] for general thin ferromagnetic shells and plates. In the case of an infinitely long thin ferromagnetic strip the formulation is simplified and becomes one-dimensional. Let $\boldsymbol{m}(t, x, y)$ be the substrate magnetization. Following [16-18], we introduce the "surface magnetization"

$$\sigma(t, x) = \int_{-\delta/2}^{\delta/2} m_x(t, x, y)\mathrm{d}y,$$

which is attributed to the substrate midsurface and satisfies the integro-differential equation

$$\sigma(t, x) + \chi\delta\partial_x\left(\frac{1}{2\pi}\int_{-a}^{a}\frac{\sigma(t, x')}{x-x'}\mathrm{d}x'\right) = \chi\delta h_x^0. \qquad (5)$$

In derivation of this equation, the boundary conditions

$$\sigma(t, \pm a) = 0 \qquad (6)$$

have been employed. Conditions (6) mean that only the magnetic flux entering through the upper and lower strip surfaces is taken into account. The flux entering through the strip edges is of the order of $\delta^2$ and can be neglected [18]. Taking (4) into account, we rewrite (5) as

$$(\chi\delta)^{-1}\sigma(t, x) + \partial_x\left(\frac{1}{2\pi}\int_{-a}^{a}\frac{\sigma(t, x')}{x-x'}\mathrm{d}x'\right) - \frac{j}{2} = h_x^{\mathrm{e}}. \qquad (7)$$

Since the current is allowed only in the superconducting layer, the following transport current condition has to be satisfied:

$$\int_{-a}^{a} j(t, x)\mathrm{d}x = I(t), \qquad (8)$$

where $I(t)$ is a known function. The field $\boldsymbol{h}^{\mathrm{fm}}$ generated by the substrate is a potential one: $\boldsymbol{h}^{\mathrm{fm}} = -\nabla\varphi^{\mathrm{fm}}$, where the scalar magnetic potential can be written (see [16, 17]) as

$$\varphi^{\mathrm{fm}}(t, x, y) = \frac{1}{4\pi}\int_{-a}^{a}\partial_{x'}\sigma(t, x')\ln\left[(x-x')^2 + y^2\right]\mathrm{d}x'.$$

Differentiating, we find

$$\boldsymbol{h}^{\mathrm{fm}} = -\frac{1}{2\pi}\begin{Bmatrix} \int_{-a}^{a}\dfrac{\partial_{x'}\sigma(t, x')(x-x')}{(x-x')^2 + y^2}\mathrm{d}x', \\ \int_{-a}^{a}\dfrac{\partial_{x'}\sigma(t, x')y}{(x-x')^2 + y^2}\mathrm{d}x' \end{Bmatrix}. \qquad (9)$$

The $x$-component of this field is continuous, while the $y$-component has a jump at $y=0, |x|<a$:

$$h_y^{\text{fm}}\Big|_{y=0-} = +\frac{\partial_x \sigma}{2}, \quad h_y^{\text{fm}}\Big|_{y=0+} = -\frac{\partial_x \sigma}{2}. \qquad (10)$$

Taking (4) and (10) into account, we find the $y$-component of the magnetic field at the superconducting layer:

$$h_y = h_y^e + \frac{1}{2\pi} \int_{-a}^{a} \frac{j(t,x')\mathrm{d}x'}{x-x'} - \frac{\partial_x \sigma}{2}. \qquad (11)$$

To complete the model, we need now to specify a constitutive relation for the superconducting layer. First, we will consider the layer in the Meissner state. In this case the normal to this layer magnetic field component, $h_y$, should be zero:

$$h_y^e + \frac{1}{2\pi} \int_{-a}^{a} \frac{j(t,x')\mathrm{d}x'}{x-x'} - \frac{\partial_x \sigma}{2} = 0. \qquad (12)$$

Second, we will consider the superconductor obeying the power current-voltage relation

$$e = e_0 \left|\frac{j}{j_c}\right|^{n-1} \frac{j}{j_c}, \qquad (13)$$

where $e(t,x)$ is the electric field, $e_0 = 10^{-4}\ \text{Vm}^{-1}$. For simplicity, in our numerical examples the power $n$ and the critical sheet current density $j_c$ will be assumed constant, however, any dependence on the magnetic field can be introduced. Instead of (12) we will now use the Faraday law and obtain

$$\partial_x e = \mu_0 \partial_t \left(h_y^e + \frac{1}{2\pi} \int_{-a}^{a} \frac{j(t,x')}{x-x'}\mathrm{d}x' - \frac{\partial_x \sigma}{2}\right). \qquad (14)$$

In this case much interest presents also the accumulating loss in the superconducting layer per unit of its length, $Q$, changing as

$$\frac{\mathrm{d}Q}{\mathrm{d}t} = \int_{-a}^{a} ej\mathrm{d}x. \qquad (15)$$

For a periodic transport current or/and external field, the loss per unit of length per period $T$ is $Q_T = Q(t+T) - Q(t)$. Here we limit our consideration to influence of a ferromagnetic substrate on AC losses in the superconducting layer; see [25, 26] for estimates of hysteretic losses in the substrate itself.

Prior to presenting our numerical scheme we would like to note the following. Superconductivity problems are, typically, solved using the finite element methods. Recently, several superconductivity problems have been solved, see [19-21], by the Chebyshev spectral methods [27], which often provide superior approximation accuracy and efficiency, although at the expense of domain flexibility. Expansions in Chebyshev polynomials simplify numerical approximation of the singular integral operators in our model. For reader's convenience, we briefly present the Chebyshev polynomials [28] and use their expansions to obtain a matrix representation of several linear operations on vectors of the mesh function values (interpolation, numerical differentiation and integration).

## III. CHEBYSHEV INTERPOLATING EXPANSIONS

The Chebyshev polynomials of the first and the second kind, $T_k(x)$ and $U_k(x)$, can be defined recursively,

$$T_0 = 1, T_1 = x,\ T_k = 2xT_{k-1} - T_{k-2}\ \text{for}\ k \geq 2,$$
$$U_0 = 1, U_1 = 2x,\ U_k = 2xU_{k-1} - U_{k-2}\ \text{for}\ k \geq 2 \qquad (16)$$

and are orthogonal on the interval [-1,1] with the weights $(1-x^2)^{-1/2}$ and $(1-x^2)^{1/2}$, respectively. We systematically use the finite expansions of two types, $f(x) = \sum_{i=0}^{N} \alpha_i T_i(x)$ and $g(x) = \sum_{i=0}^{N} \beta_i U_i(x)$. Let $x_0, ..., x_N$ be a mesh in [-1,1]. Provided the vectors of expansion coefficients $\overline{\alpha}$ and $\overline{\beta}$ are given, vectors of the mesh values $f_i = f(x_i)$ and $g_i = g(x_i)$ are $\overline{f} = P\overline{\alpha}$ and $\overline{g} = R\overline{\beta}$, where the matrix elements $P_{ij} = T_j(x_i)$ and $R_{ij} = U_j(x_i)$ can be efficiently calculated using the recursive relations (16). If, on the contrary, the mesh values of functions $f$ or $g$ are given, the Chebyshev expansions interpolating these values have the coefficients $\overline{\alpha} = P^{-1}\overline{f}$ and $\overline{\beta} = R^{-1}\overline{g}$. Please note that we numerate rows and columns in all matrices starting from zero, as this better complies with the numeration of Chebyshev polynomials.

As is well known, even for a smooth function its polynomial interpolant on a uniform mesh can demonstrate violent fluctuations between the interpolation knots (the Runge phenomenon). Such undesirable behavior is efficiently suppressed by using non-uniform meshes with the interpolation knots that are placed denser near the interval ends. Two such meshes are typically employed with Chebyshev interpolating expansions: the Chebyshev points of the first kind, not including the interval ends $x = \pm 1$,

$$x_i = -\cos\left[\pi \frac{i+1/2}{N+1}\right], \quad i = 0, ..., N, \qquad (17)$$

and of the second kind, which include the end points,

$$x_i = -\cos\left[\pi \frac{i}{N}\right], \quad i = 0, ..., N. \qquad (18)$$

Below, we will use both these meshes.

Let the expansion $\sum_{k=0}^{N} \alpha_k T_k(x)$ interpolates the mesh values $f_i = f(x_i)$. Then $\overline{\alpha} = P^{-1}\overline{f}$ and, since

$$\mathrm{d}T_k(x)/\mathrm{d}x = kU_{k-1}(x),$$

we set $\mathrm{d}f/\mathrm{d}x \approx \sum_{k=0}^{N} k\alpha_k U_{k-1}(x)$. Introducing the $(N+1)\times(N+1)$ sparse matrix with the non-zero elements $C_{k,k+1} = k+1$, $k = 0,...,N-1$, we can write the coefficients of this expansion as $\overline{\beta} = CP^{-1}\overline{f}$. Its mesh values, approximating the mesh values of the derivative, are $\overline{f}' = D\overline{f}$, where

$$D = RCP^{-1} \qquad (19)$$

is the Chebyshev differentiation matrix.

To integrate a function given by its mesh values, we interpolate $f(x)\sqrt{1-x^2}$ by $\sum_{k=0}^{N}\gamma_k T_k(x)$. Then $\overline{\gamma} = P^{-1}X\overline{f}$, where $X$ is the diagonal matrix with $X_{ii} = \sqrt{1-x_i^2}$. Since $\int_{-1}^{1} T_k(x)(1-x^2)^{-1/2}\mathrm{d}x$ is $\pi$ for $k = 0$ and zero otherwise, we have

$$\int_{-1}^{1} f(x)\mathrm{d}x \approx \int_{-1}^{1}\frac{\sum_{k=0}^{N}\gamma_k T_k(x)}{\sqrt{1-x^2}}\mathrm{d}x = \pi\gamma_0 .$$

Hence, our quadrature formula is

$$\int_{-1}^{1} f(x)\mathrm{d}x \approx \overline{w}\cdot\overline{f}, \qquad (20)$$

where $\overline{w}$ is the first row of $\pi P^{-1}X$.

To deal with the singular integral operators in our model, we use the well-known equality: for $x\in(-1,1)$ and $k\geq 0$

$$\int_{-1}^{1}\frac{T_k(x')\mathrm{d}x'}{(x-x')\sqrt{1-(x')^2}} = -\pi U_{k-1}(x),$$

where $U_{-1} = 0$. Interpolating $f(x)\sqrt{1-x^2}$ in the mesh points by $\sum_{k=0}^{N}\gamma_k T_k(x)$ again, we obtain

$$J^f(x) := \frac{1}{2\pi}\int_{-1}^{1}\frac{f(x')\mathrm{d}x'}{x-x'} \approx -\frac{1}{2}\sum_{k=1}^{N}\gamma_k U_{k-1}(x)$$

with $\overline{\gamma} = P^{-1}X\overline{f}$. Denoting $F_{ij} = -1/2$ if $j = i+1$ and zero otherwise, we set for the mesh values

$$\overline{J}^f = S\overline{f}, \qquad (21)$$

where $S = RFP^{-1}X$.

## IV. Superconductor in the Meissner state

We use the scaling

$$\tilde{x} = x/a, \quad \tilde{y} = y/a, \quad \tilde{\sigma} = \sigma/a, \quad \tilde{I} = I/a$$

and, omitting the sign "$\sim$", rewrite the equations (7), (12) and (8) in the scaled variables for each moment in time:

$$\frac{a}{\chi\delta}\sigma(x) + \partial_x\left(\frac{1}{2\pi}\int_{-1}^{1}\frac{\sigma(x')}{x-x'}\mathrm{d}x'\right) - \frac{j}{2} = h_x^e, \qquad (22)$$

$$h_y^e + \frac{1}{2\pi}\int_{-1}^{1}\frac{j(x')\mathrm{d}x'}{x-x'} - \frac{\partial_x\sigma}{2} = 0, \qquad (23)$$

$$\int_{-1}^{1} j(x)\mathrm{d}x = I. \qquad (24)$$

In the Meissner state model of a thin superconducting film, the sheet current density $j$ is infinite at the film edges. This is why in this case we choose the Chebyshev points of the first kind, $x_0,...,x_N$, as a mesh not including the end points $x = \pm 1$. Employing (19)-(21), we discretize the equations (22)-(24) and obtain

$$\left(\kappa^{-1}E + DS\right)\overline{\sigma} - \frac{\overline{j}}{2} = \overline{h}_x^e, \qquad (25)$$

$$\frac{1}{2}D\overline{\sigma} - S\overline{j} = \overline{h}_y^e, \qquad (26)$$

$$\overline{w}\cdot\overline{j} = I, \qquad (27)$$

where $\kappa = \chi\delta/a$ and $E$ is the unit matrix. The system consists of $2N+3$ linear algebraic equations for $2N+2$ coordinates of the vectors $\overline{\sigma}$ and $\overline{j}$. If, however, we multiply (26) by the full rank matrix $R^{-1}$, the last of the equivalent to (26) set of linear equations,

$$\frac{1}{2}R^{-1}D\overline{\sigma} - R^{-1}S\overline{j} = R^{-1}\overline{h}_y^e, \qquad (28)$$

is the trivial $0 = 0$. Indeed, the last rows of matrices $R^{-1}D = CP^{-1}$ and $R^{-1}S = FP^{-1}X$ are zeros, since such are the last rows of $C$ and $F$. On the right, $R^{-1}\overline{h}_y^e$ is the vector of coefficients of the interpolating expansion $h_y^e \approx \sum_{k=0}^{N}\beta_k U_k(x)$, so the right hand side of the last equation is $\beta_N$. This coefficient is zero if $h_y^e$ is spatially uniform and is vanishingly small for any smooth function. In the latter case we can use the approximation of $h_y^e$ with $\beta_N = 0$. This allows us to find $\overline{\sigma}$ and $\overline{j}$ by solving the linear system consisting of (25) and also (28) with the last equation replaced by (27).

Magnetic field outside the coated conductor is defined by (1), (2) and (9). In the scaled variables

$$h_x = h_x^e - \frac{1}{2\pi} \int_{-1}^{1} \frac{j(t,x')y + \partial_{x'}\sigma(t,x')(x-x')}{(x-x')^2 + y^2} \, dx',$$

$$h_y = h_y^e + \frac{1}{2\pi} \int_{-1}^{1} \frac{j(t,x')(x-x') - \partial_{x'}\sigma(t,x')y}{(x-x')^2 + y^2} \, dx'. \tag{29}$$

Provided that $\bar{\sigma}$ and $\bar{j}$ are found, we compute this field numerically using (19) and (20):

$$h_x(x,y) = h_x^e(x,y) - \frac{1}{2\pi} \sum_{i=0}^{N} w_i \frac{j_i y + \psi_i(x-x_i)}{(x-x_i)^2 + y^2},$$

$$h_y(x,y) = h_y^e(x,y) + \frac{1}{2\pi} \sum_{i=0}^{N} w_i \frac{j_i(x-x_i) - \psi_i y}{(x-x_i)^2 + y^2}, \tag{30}$$

where $\bar{\psi} = D\bar{\sigma}$. Results of three numerical simulations for $\kappa = 5$ and $N = 200$ are presented in Figs 1-3. The boundary conditions $\sigma(\pm 1) = 0$ were not directly used in these computations; however, plotting $\sigma$ we added the end points for clarity.

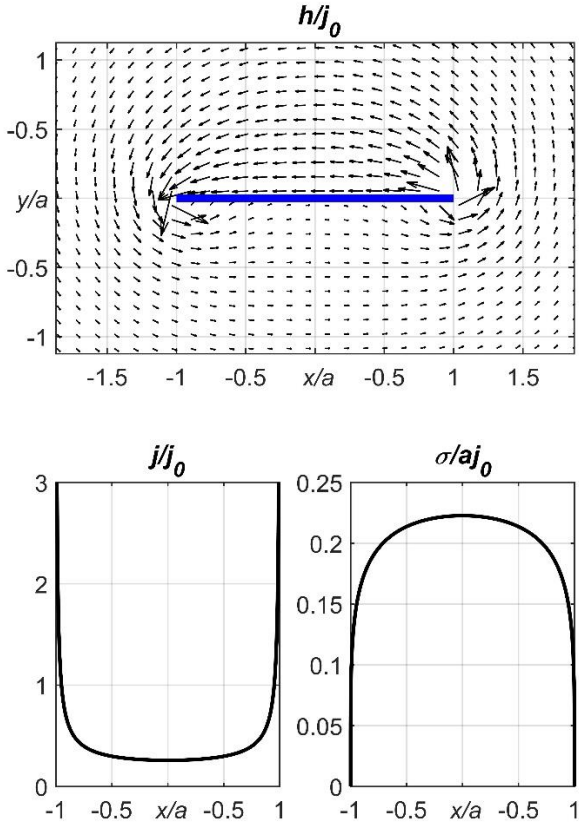

**Fig. 1.** Meissner state, $\boldsymbol{h}^e = 0$, $I \neq 0$. Top - magnetic field around the coated conductor. Bottom - the sheet current density $j$ (left) and surface magnetization $\sigma$ (right). Scaling coefficient $j_0 = I / a$.

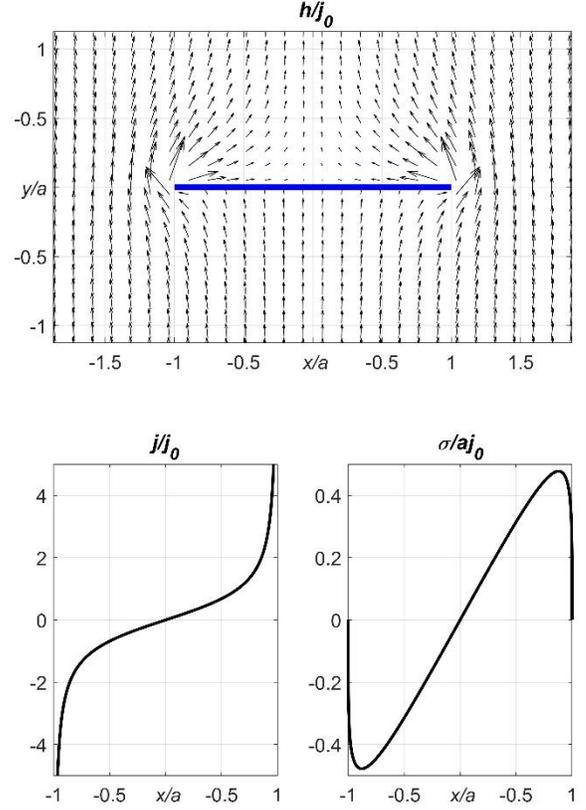

**Fig. 2.** As in Fig. 1 but for $h_y^e \neq 0, h_x^e = 0, I = 0$. Scaling coefficient $j_0 = h_y^e$.

The model (22)-(24) contains a single parameter $\kappa = \chi\delta / a$. For $\kappa \ll 1$ equation (22) yields $\sigma \to 0$ and the influence of the magnetic substrate is negligible; this is the situation considered in [24]. The case $\kappa \gg 1$ corresponds to an ideal substrate with $\chi \to \infty$ as was assumed in [1]; in this case the $x$-component of the total magnetic field just below the substrate is zero. Using (6) and (9), we can write this as

$$h_x^e - \frac{1}{2\pi} \int_{-1}^{1} \frac{\partial_{x'}\sigma(x')}{x-x'} \, dx' + \frac{j}{2} = 0, \tag{31}$$

which should be solved together with (23) and (24).

As an example, we consider the influence of $\kappa$ on the sheet current density in a transport current problem and compare our solution to the analytical ones in the two limiting cases. The Brandt and Indenbom solution [24] is $j = I / \left(\pi\sqrt{a^2 - x^2}\right)$. For the Mawatari solution we use here an explicit formula for $j$ that is, probably, more convenient for calculations than the indirect one, via the magnetic field, in [1].

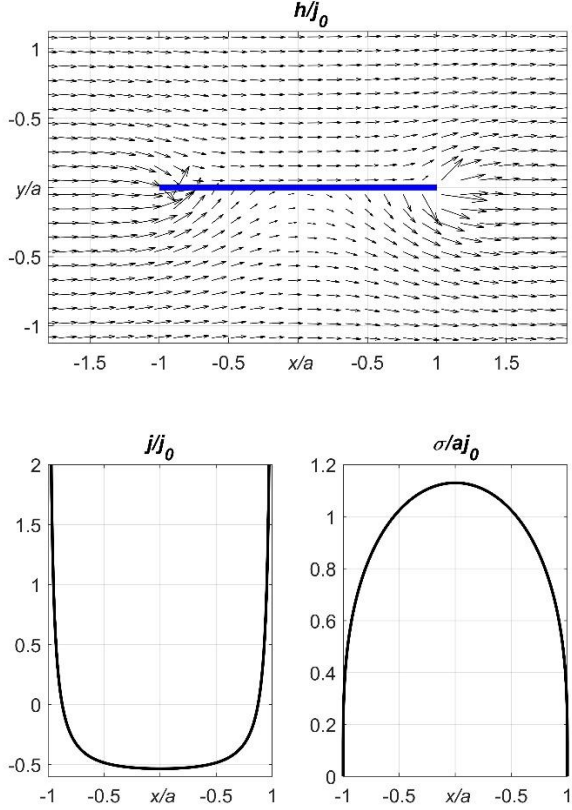

**Fig. 3.** As in Fig. 1 but for $h_x^e \neq 0$, $h_y^e = 0$, $I = 0$. Scaling coefficient $j_0 = h_x^e$.

Denoting

$$u = j + \partial_x \sigma, \quad v = j - \partial_x \sigma$$

we obtain, using (23) and (31), singular integral equations for each of these variables alone:

$$v + \frac{1}{\pi} \int_{-1}^{1} \frac{v(x')}{x - x'} dx' = -2(h_x^e + h_y^e),$$

$$u - \frac{1}{\pi} \int_{-1}^{1} \frac{u(x')}{x - x'} dx' = 2(h_y^e - h_x^e).$$

These equations can be solved analytically ([29], par. 4.1-7, N66.3) and, for the transport current problem with $h_y^e = h_x^e = 0$ in our example, we have

$$j = (u + v) / 2 =$$
$$C \left[ (1+x)^{-1/4}(1-x)^{-3/4} + (1+x)^{-3/4}(1-x)^{-1/4} \right],$$

where $C$ is an arbitrary constant. This constant is determined by (24) and, integrating (see [30], 3.196, n.3), we find $C = [2B(1/4, 3/4)]^{-1} I = (2\pi\sqrt{2})^{-1} I$, where $B$ is the beta function. Computations with different values of

$\kappa$ (Fig. 4) confirm that for small $\kappa$ our solution is close to the analytical solution for a superconducting strip with a nonmagnetic substrate [24] and, for large $\kappa$, to the Mawatari solution [1] for the substrate with an infinite permeability.

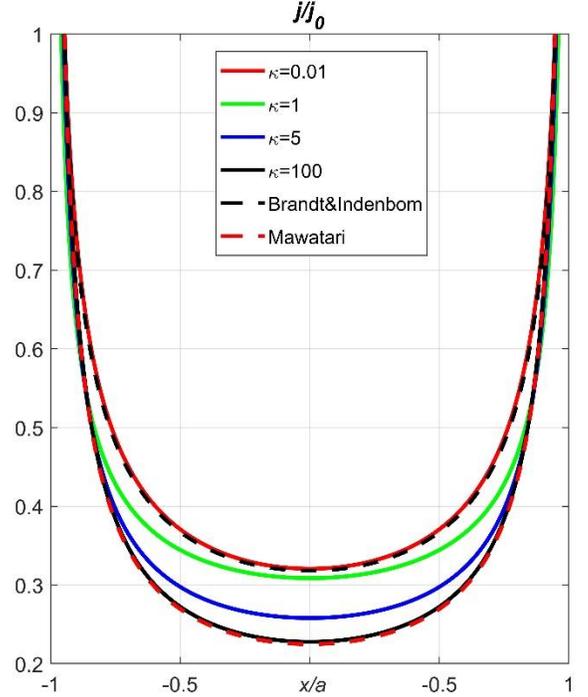

**Fig. 4.** Meissner state, solution of the transport current problem with different values of $\kappa$ and the analytical solutions [1] (red dashed curve) and [24] (black dashed curve). Scaling coefficient $j_0 = I / a$.

## V. Superconductor with a Nonlinear Current-Voltage Relation

We now assume for the superconductor the power $e(j)$ relation (13) with constant parameters $j_c$ and $n$. It is convenient to use the scaled variables

$$\tilde{j} = \frac{j}{j_c}, \quad \tilde{\sigma} = \frac{\sigma}{a j_c}, \quad \tilde{h} = \frac{h}{j_c}, \quad \tilde{e} = \frac{e}{e_0},$$

$$(\tilde{x}, \tilde{y}) = \frac{(x, y)}{a}, \quad \tilde{t} = \frac{t}{t_0}, \quad \tilde{I} = \frac{I}{I_c}, \quad \tilde{Q} = \frac{Q}{Q_0},$$

where $t_0 = a\mu_0 j_c / e_0$, the critical current $I_c = 2aj_c$, and, as in [31], $Q_0 = \mu_0 I_c^2$. Omitting the sign "~" to simplify the notations, we rewrite our model equations (7),(8),(13) and (14) in dimensionless form:

$$\kappa^{-1}\sigma(t,x)+$$
$$\partial_x\left(\frac{1}{2\pi}\int_{-1}^{1}\frac{\sigma(t,x')}{x-x'}dx'\right)-\frac{j(t,x)}{2}=h_x^e(t,x), \tag{32}$$

$$\int_{-1}^{1}j(t,x)dx=2I(t), \tag{33}$$

$$e=|j|^{n-1}j, \tag{34}$$

$$\partial_x e(t,x)=$$
$$\partial_t\left(h_y^e+\frac{1}{2\pi}\int_{-1}^{1}\frac{j(t,x')}{x-x'}dx'-\frac{\partial_x\sigma(t,x)}{2}\right). \tag{35}$$

In this model the sheet current density $j$ remains bounded at the strip edges and we can use the Chebyshev mesh of the second kind (18). Moreover, we found that if the mesh does not include the end points, numerical computations become unstable.

The problem (32)-(35) is evolutionary and for its integration in time we use the method of lines. Knowing the mesh values $\overline{j}(t)$ and $\overline{\sigma}(t)$ at time $t$, to proceed further we need to compute their derivatives, $\dot{\overline{j}}(t)$ and $\dot{\overline{\sigma}}(t)$ (here and below the dot above a variable means the time derivative). Using (19)-(21), we obtain the spatially discretized version of (32)-(35):

$$\left(\kappa^{-1}E+DS\right)\overline{\sigma}-\frac{\overline{j}}{2}=\overline{h}_x^e, \tag{36}$$

$$\overline{w}\cdot\overline{j}=2I, \tag{37}$$

$$\overline{e}=|\overline{j}|^{n-1}\overline{j}, \tag{38}$$

$$D\overline{e}=\dot{\overline{h}}_y^e+S\dot{\overline{j}}-\frac{1}{2}D\dot{\overline{\sigma}}. \tag{39}$$

The discretized dimensionless form of (15) is

$$\dot{Q}=\frac{1}{4}\sum_{k=0}^{N}w_k j_k e_k. \tag{40}$$

Let us differentiate (36) in time and multiply (39) by the full rank matrix $R^{-1}$. This yeilds

$$\left(\kappa^{-1}E+DS\right)\dot{\overline{\sigma}}-\frac{\dot{\overline{j}}}{2}=\dot{\overline{h}}_x^e, \tag{41}$$

$$R^{-1}S\dot{\overline{j}}-\frac{1}{2}R^{-1}D\dot{\overline{\sigma}}=R^{-1}D\overline{e}-R^{-1}\dot{\overline{h}}_y^e. \tag{42}$$

As in the Meissner state case, the last of equations (42) is trivial, $0=0$, since the last rows of matrices $R^{-1}S$ and $R^{-1}D$ are zeros and on the right hand side of this equation there remains only $-\beta_N$, the last coefficient in the interpolating Chebyshev expansion $\dot{h}_y^e\approx\sum_{i=0}^{N}\beta_i U_i(x)$; this coefficient can be always set to zero. Knowing $\overline{j}$ at time $t$, we compute $\overline{e}$ using (38) and can find the time derivatives $\dot{\overline{j}}(t)$ and $\dot{\overline{\sigma}}(t)$ by solving the linear system, consisting of (41) and (42) in which the last equation is replaced by the time derivative of (37), $\overline{w}\cdot\dot{\overline{j}}=2dI/dt$.

Numerical simulations showed, however, that the obtained solution is inaccurate at the two end points, $x_0$ and $x_N$, and conditions (6) are violated. Solution of the magnetostatic problem for a thin plate depends, near the edges, on the edge shape. Without entering the detailed analysis (see [16, 32]), we resolved this complication by reducing the thickness of the substrate at the end points to almost zero (to $10^{-6}\delta$ in our simulations) with the decrease of the $\kappa$ values at these points, and replacing $\kappa^{-1}E$ in (41) by the diagonal matrix $\mathrm{diag}(\overline{\kappa})^{-1}$.

Denoting the $(2N+2)\times(2N+2)$ matrix of thus transformed system (41)-(42) by $A$ and the $(2N+2)$-vector of the right hand sides by $\overline{r}$, we notice that the matrix $A$ is constant and needs to be inverted only once. The system of $2N+2$ ordinary differential equations (ODE) to be solved is

$$\frac{d}{dt}\begin{bmatrix}\overline{j}\\\overline{\sigma}\end{bmatrix}=A^{-1}\overline{r}(t,\overline{j}).$$

To follow the accumulating losses, this system was supplemented by the ODE equation (40). Finally, the system was integrated using the Matlab ODE solver ode15s. Equations (30) were used also in this case for computing the magnetic field around the coated conductor.

Convergence rate of this method was investigated for $n=30$ and the transport current growing linearly with time as $I=20t$ in dimensionless variables. Solution for $I=0.75$ was found using the Chebyshev meshes of the second kind with $N=25,50,100,200,400$ and also $N_0=800$; both the relative and the absolute tolerances of the ODE solver were set to $10^{-8}$. For each $N$ we constructed the Chebyshev interpolating expansions, $j^N(x)$ and $\sigma^N(x)$, of the numerical solutions $\overline{j}^N$ and $\overline{\sigma}^N$, respectively. Then found the relative deviations (in the integral $L^1$-norm) from the most accurate solution, $\overline{j}^{N_0}$ and $\overline{\sigma}^{N_0}$, in the nodes $x_k^{N_0}$:

$$\delta^N(j)=\frac{\sum_{k=0}^{N_0}w_k^{N_0}\left|j^N(x_k^{N_0})-j_k^{N_0}\right|}{\sum_{k=0}^{N_0}w_k^{N_0}\left|j_k^{N_0}\right|},$$

$$\delta^N(\sigma)=\frac{\sum_{k=0}^{N_0}w_k^{N_0}\left|\sigma^N(x_k^{N_0})-\sigma_k^{N_0}\right|}{\sum_{k=0}^{N_0}w_k^{N_0}\left|\sigma_k^{N_0}\right|},$$

where $w_k^{N_o}$ are the quadrature coefficients. These relative error estimates and also the computation times, approximately determined by Matlab, are presented in Table I and illustrate the efficiency and accuracy of this method. Even for $N=25$ the errors are less than 1%.

TABLE I.
CONVERGENCE RATE AND COMPUTATION TIMES

| $N$ | $\delta^N(j)$ | $\delta^N(\sigma)$ | CPU time (seconds) |
|---|---|---|---|
| 25 | 9.9e-3 | 3.8e-3 | 0.16 |
| 50 | 4.0e-3 | 1.3e-3 | 0.17 |
| 100 | 1.4e-3 | 2.9e-4 | 0.47 |
| 200 | 1.2e-4 | 7.5e-5 | 2.4 |
| 400 | 1.3e-5 | 1.7e-5 | 15 |
| 800 | - | - | 112 |

Numerical simulation results in Figs 5-7 were obtained for $n=30$, $\kappa=5$ and $N=200$.

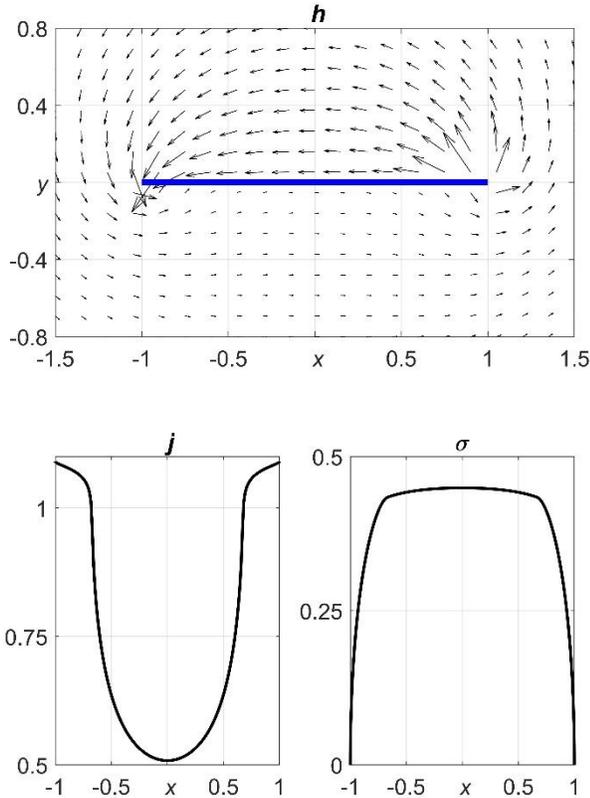

**Fig. 5.** Power $e(j)$ relation, $I=0.75$, $\boldsymbol{h}^e=0$. Top – magnetic field; bottom – sheet current density (left) and surface magnetization (right). Scaled dimensionless variables.

In these examples we assumed that either the transport current $I$ or one of the external field components ($h_x^e$ or $h_y^e$) is increased linearly with time as $20t$.

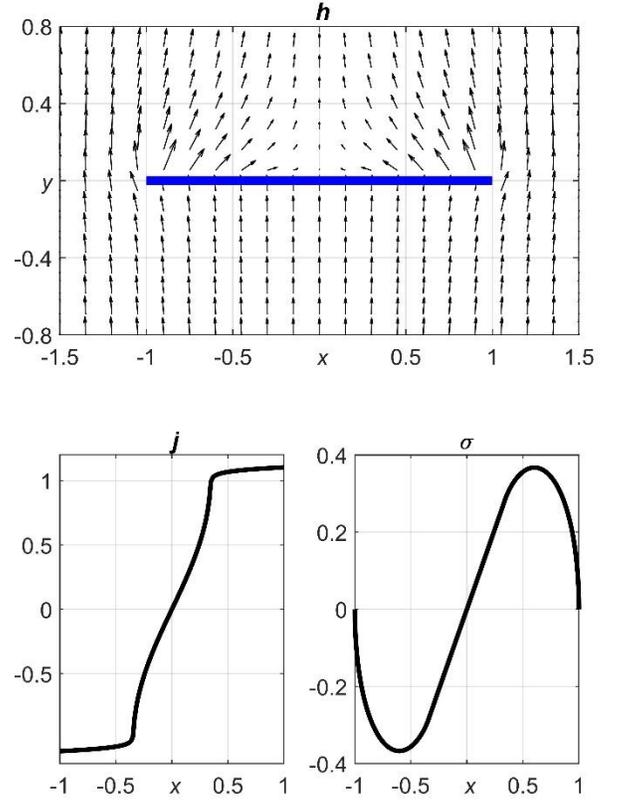

**Fig. 6.** As in f Fig. 5 but for $h_y^e=1$, $h_x^e=0$, $I=0$.

Mathematically, convergence of the power law model for a superconductor to the Bean model as $n \to \infty$ was shown in [33]. Choosing the high power $n=200$ and different values of $\kappa$, we compared our model solutions for the normal external field to those derived by Brandt and Indenbom [24] and by Mawatari [1] for superconductors described by the critical-state model (Fig. 8). A Matlab code for calculating the analytical solution [1] was in this case kindly provided by Dr Cun Xue [34]. As in the Meissner case, our solution is close to that in [24] for small $\kappa$ and to the solution in [1] for large $\kappa$.

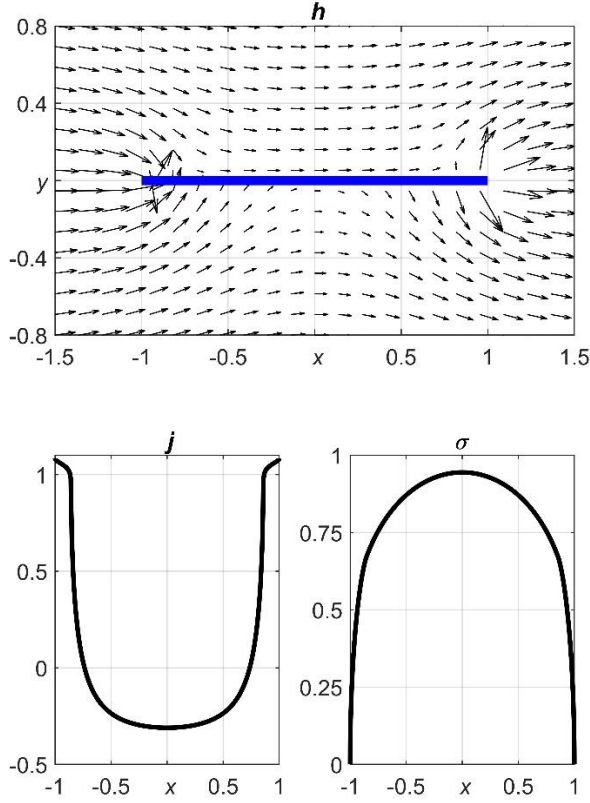

**Fig. 7.** As in Fig. 5 but for $h_x^e = 1$, $h_y^e = 0$, $I = 0$.

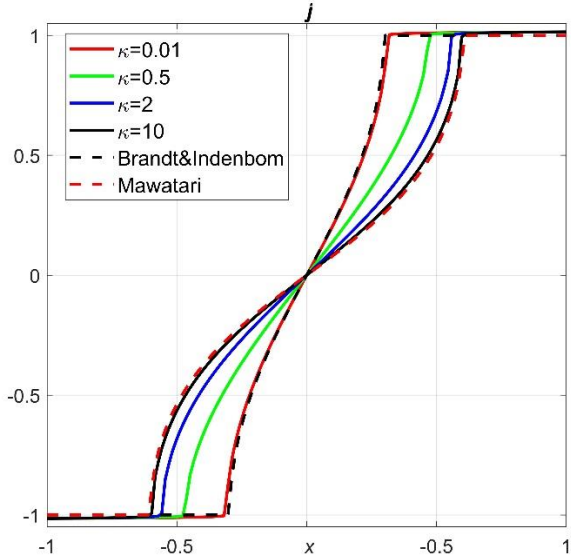

**Fig. 8.** Sheet current density (dimensionless) for different values of $\kappa$. Power $e(j)$ relation with $n = 200$ for comparison with the critical-state solutions in [1] and [24]; $h_y^e = 0.6$, $h_x^e = 0$, $I = 0$.

Assuming $\mathbf{h}_y^e = \left\{ H_y \sin(2\pi t / T), 0 \right\}$ with $T = 0.025$, we solved the problem for $0 \le t \le 2T$ and different amplitudes $H_y$ to estimate the second cycle losses $Q_T = Q_T(H_y)$ for a coated conductor in a periodic perpendicular field. For, e.g., $a = 5$ mm, and the critical sheet current density $j_c \approx 3 \cdot 10^4 \mathrm{Am}^{-1}$ in [31], the time scaling factor $t_0 \approx 0.5$ s and our choice of $T$ corresponds to the frequency 20 Hz. The losses, computed for $n = 30$ and different values of $\kappa$, are presented as a log-log plots of $Q_T(H_y) / H_y^2$ in Fig. 9. For $\kappa = 0.01$ (Fig. 9, dashed curve) the losses should be close to those for a non-magnetic substrate. For fields, having the amplitudes $H_y$ lower than some, depending on $\kappa$, value not exceeding 0.14, the losses are significantly higher if the substrate is magnetic. Then the curves intersect and, for higher field amplitudes, the losses for the magnetic substrate case are lower. The difference, however, becomes negligible as the field amplitude grows and exceeds 5. For $\kappa = 3$ the loss curve is already not too far from the $\kappa = 100$ curve, which should be close to the $\kappa = \infty$ curve. This suggests that, for e.g., $\kappa > 5$ Mawatari's assumption $\chi = \infty$ can be used to accurately predict the losses in oscillating fields. This should be true also for the field-dependent susceptibility of the substrate provided that $\kappa$ remains sufficiently large. Our results are in a qualitative agreement with the experimental data [31], where the effective substrate susceptibility was estimated as 30; with $a = 5$ mm and $\delta = 100$ μm this gives $\kappa \approx 0.6$. The magnetic and non-magnetic substrate loss curves in that work also intersect at an $H_y$ between 0.1 and 0.2. The maximal $Q_T(H_y) / H_y^2$ value, reported in that work, $0.34$, was reached for $H_y \approx 0.9$. Our simulation with this $\kappa$ yielded 0.49 and 1.1, respectively. The difference can result from the simplifying assumptions in our model, such as the constant (field-independent) values of $\chi$ and $j_c$. We note also that the superconducting layer and substrate in [31] were separated by a $25$ μm buffer layer absent in our model (the buffer layer in coated conductors is usually less than 1 μm thick). For $n = 100$ the power $e(j)$ law model is an approximation to the Bean critical-state model with frequency-independent losses; our results for small and large $\kappa$ (Fig. 9, inset) are close to losses found for the critical-state model analytically in [24] and [1, 5], respectively.

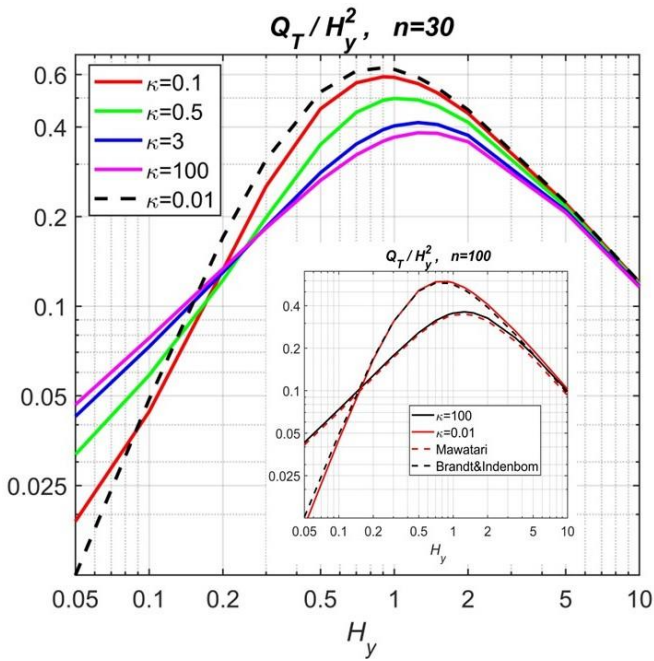

**Fig. 9.** Hysteretic losses in the superconductor; $n = 30$, AC perpendicular field $h_y^e = H_y \sin(2\pi t / T)$ with $T = 0.025$. Inset: results for $n = 100$ and the analytical solutions [1, 5] and [24]. Dimensionless variables.

For the periodic transport current $I = I_0 \sin(2\pi t / T)$ the scaled loss $Q_T / I_0^2$ (Fig. 10) increases monotonically with transport current amplitude $I_0$ and also with the value of $\kappa$; in these simulations we set again $T = 0.025$ and $n = 30$. Using $n = 100$ to simulate the critical-state model behavior, we checked that for small $\kappa$ the loss tends to that calculated analytically for non-magnetic substrates by Norris [35] and for large $\kappa$ to Mawatari's solution [1, 5] obtained for $\chi = \infty$ (Fig. 10, inset).

The loss in a parallel AC field is negligibly small if the substrate is non-magnetic (zero in the infinitely thin superconductor model). Magnetic substrates radically change the situation: they modify the magnetic field direction and a normal field component appears. The losses (Fig. 11) first increase quickly with $\kappa$, then tend to a limiting value as $\kappa \to \infty$, and are always significantly lower than the losses in a perpendicular AC field of the same amplitude. In this case no analytical expression for losses is known to us.

Although our model assumes constant susceptibility and sheet critical current density, qualitatively, presented results for the AC current and applied field agree well with experiments and previous simulations [13-15, 36].

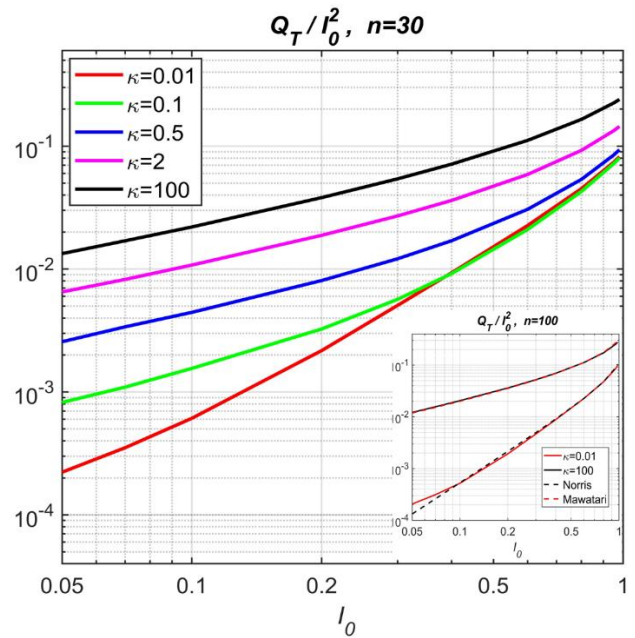

**Fig. 10.** As in Fig. 9 but for the AC transport current $h_y^e = H_y \sin(2\pi t / T)$. Inset: results for $n = 100$ and the analytical solutions [35] and [1, 5].

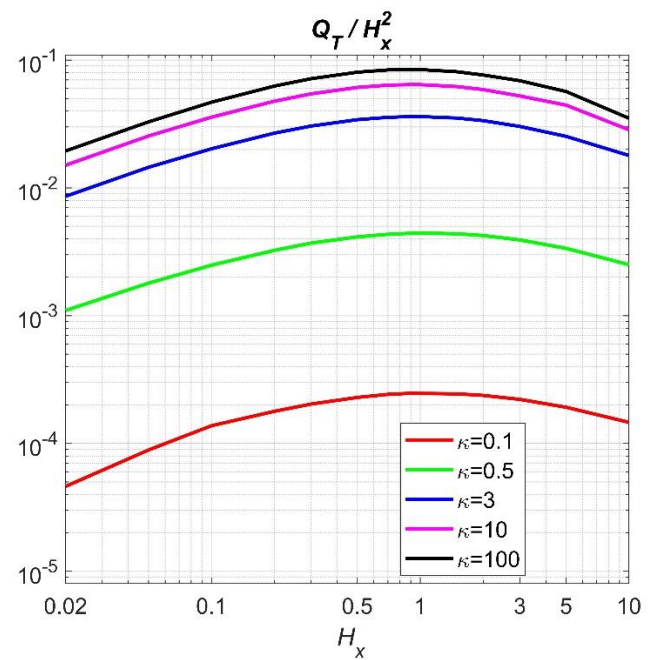

**Fig. 11.** As in Fig. 9 but for the AC parallel external field $h_x^e = H_x \sin(2\pi t / T)$; $n = 30$.

## VI. Conclusion

While mathematical models and numerical methods for modeling thin superconducting films are well developed, the problem becomes significantly more complicated, even for the simplest geometry of an infinitely long coated conductor, if it is necessary to account for a ferromagnetic substrate. These substrates change the distribution of superconducting currents and influence the AC losses in the superconducting layer. Numerical simulations of such bilayer systems in previous works did not take into account that, usually, the thickness of a substrate is small comparing to its other geometrical sizes. By using the quasistatic thin shell magnetization model developed by Krasnov [16-18], we could drastically simplify the problem formulation and present it as a system of one-dimensional singular integro-differential equations for the sheet current density in the superconducting layer and the effective (surface) magnetization of the substrate. Only one dimensionless parameter, $\kappa = \chi \delta / a$, characterizes the magnetic substrate in this model. Our numerical method, employing the Chebyshev polynomial expansions for spatial approximation and the method of lines for integration in time, is fast and accurate; this, in particular, is achieved due to the analytical treatment of the singular integrals.

The aim of our work is to present the new model and numerical scheme. For simplicity, we assumed a constant susceptibility of the substrate material and a constant critical sheet current density in the power current-voltage relation for the superconductor. Numerical simulations and comparison with experimental and numerical results in other works showed that the model correctly describes the main features of electromagnetic response of a coated conductor with magnetic substrate to variations of the transport current and applied field. In the limiting cases of weak and strong magnetic substrates (i.e. with small and large $\kappa$) our numerical solution tends to the known analytical solutions. It is straightforward to replace the power current-voltage relation for a superconductor by any other, including a field-dependent one. In future, we are going to extend the model to problems with a field-dependent magnetic susceptibility of a substrate and also to problems with arbitrary shaped superconducting films on a magnetic substrate.

## References

[1] Y. Mawatari, "Magnetic field distributions around superconducting strips on ferromagnetic substrates," *Physical Review B,* vol. 77, no. 10, Art. no. 104505, 2008.

[2] Y. Genenko, H. Rauh, and P. Krüger, "Finite-element simulations of hysteretic ac losses in a bilayer superconductor/ferromagnet heterostructure subject to an oscillating transverse magnetic field," *Applied Physics Letters,* vol. 98, no. 15, Art. no. 152508, 2011.

[3] G.-T. Ma, "Hysteretic ac loss in a coated superconductor subjected to oscillating magnetic fields: Ferromagnetic effect and frequency dependence," *Superconductor Science and Technology,* vol. 27, no. 6, Art. no. 065011, 2014.

[4] S. Li, D.-X. Chen, and J. Fang, "Transport ac losses of a second-generation HTS tape with a ferromagnetic substrate and conducting stabilizer," *Superconductor Science and Technology,* vol. 28, no. 12, Art. no. 125011, 2015.

[5] G. P. Mikitik, Y. Mawatari, A. T. Wan, and F. Sirois, "Analytical methods and formulas for modeling high temperature superconductors," *IEEE Transactions on Applied Superconductivity,* vol. 23, no. 2, Art. no. 8001920, 2013.

[6] J. Hu *et al.,* "Impact of magnetic substrate on dynamic loss and magnetization loss of HTS coated conductors," *IEEE Transactions on Applied Superconductivity,* vol. 32, no. 4, Art. no. 8200405, 2022.

[7] Y. Statra, H. Menana, and B. Douine, "Integral Modeling of AC Losses in HTS Tapes With Magnetic Substrates," *IEEE Transactions on Applied Superconductivity,* vol. 32, no. 2, Art. no. 5900407, 2021.

[8] G. T. Ma and H. Rauh, "Thermo-electromagnetic properties of a magnetically shielded superconductor strip: theoretical foundations and numerical simulations," *Superconductor Science and Technology,* vol. 26, no. 10, Art. no. 105001, 2013.

[9] F. Gomory, E. Pardo, M. Vojenciak, and J. Souc, "Magnetic flux penetration and transport AC loss in superconductor coated conductor on ferromagnetic substrate," *IEEE transactions on applied superconductivity,* vol. 19, no. 3, pp. 3102-3105, 2009.

[10] G. Liu *et al.,* "Influence of substrate magnetism on frequency-dependent transport loss in HTS-coated conductors," *IEEE Transactions on Applied Superconductivity,* vol. 27, no. 8, Art. no. 015010, 2017.

[11] X. Wan, C. Huang, H. Yong, and Y. Zhou, "Effect of the magnetic material on AC losses in HTS conductors in AC magnetic field carrying AC transport current," *AIP Advances,* vol. 5, no. 11, Art. no. 117139, 2015.

[12] M. Umabuchi, D. Miyagi, N. Takahashi, and O. Tsukamoto, "Analysis of AC loss properties of HTS coated-conductor with magnetic substrate under external magnetic field using FEM," *Physica C: Superconductivity,* vol. 468, no. 15-20, pp. 1739-1742, 2008.

[13] L. Jiang, C. Xue, and Y.-H. Zhou, "Parallel magnetic field dependence of AC losses in periodically arranged superconductors with ferromagnetic substrates," *Physica C: Superconductivity and its Applications,* vol. 566, Art. no. 1353521, 2019.

[14] D. Miyagi, M. Umabuchi, N. Takahashi, and O. Tsukamoto, "FEM analysis of effect of nonlinear magnetic property of substrate on magnetization loss in HTS layer of coated conductor," *IEEE Transactions on Applied Superconductivity,* vol. 18, no. 2, pp. 1374-1377, 2008.

[15] D. Miyagi, M. Umabuchi, N. Takahashi, and O. Tsukamoto, "Numerical Evaluation of AC

Magnetization Loss Characteristics of HTS Coated Conductor With Magnetic Substrate Using FEM," *IEEE Transactions on Applied Superconductivity,* vol. 19, no. 3, pp. 3336-3339, 2009.

[16] I. P. Krasnov, "Integral equation for magnetostatic problems with thin plates or shells," *Soviet Physics. Technical Physics,* vol. 22, no. 7, pp. 811-817, 1977.

[17] I. P. Krasnov, "Solution of the magnetostatic equations for thin plates and shells in the slab and axisymmetric cases," *Soviet physics. Technical physics,* vol. 27, no. 5, pp. 535-538, 1982.

[18] I. P. Krasnov, *Computational Methods of Ship Magnetism and Electric Engineering* [in Russian]. Leningrad: Sudostroenie 1986.

[19] V. Sokolovsky, L. Prigozhin, and A. B. Kozyrev, "Chebyshev spectral method for superconductivity problems," *Superconductor Science and Technology,* vol. 33, no. 8, Art. no. 085008, 2020.

[20] L. Prigozhin and V. Sokolovsky, "Fast solution of the superconducting dynamo benchmark problem," *Superconductor Science and Technology,* vol. 34, no. 6, Art. no. 065006, 2021.

[21] V. Sokolovsky and L. Prigozhin, "Hermite-Chebyshev pseudospectral method for inhomogeneous superconducting strip problems and magnetic flux pump modeling," *Superconductor Science and Technology,* vol. 35, no. 2, Art. no. 024002, 2021.

[22] D. Miyagi, Y. Yunoki, M. Umabuchi, N. Takahashi, and O. Tsukamoto, "Measurement of magnetic properties of Ni-alloy substrate of HTS coated conductor in LN2," *Physica C: Superconductivity,* vol. 468, no. 15, pp. 1743-1746, 2008.

[23] M. Halse, "AC face field losses in a type II superconductor," *Journal of Physics D: Applied Physics,* vol. 3, no. 5, pp. 717-720, 1970.

[24] E. H. Brandt and M. Indenbom, "Type-II-superconductor strip with current in a perpendicular magnetic field," *Physical review B,* vol. 48, no. 17, p. 12893-12905, 1993.

[25] F. Gömöry, M. Vojenčiak, E. Pardo, M. Solovyov, and J. Šouc, "AC losses in coated conductors," *Superconductor Science and Technology,* vol. 23, no. 3, Art. no. 034012, 2010.

[26] D. N. Nguyen, S. P. Ashworth, J. O. Willis, F. Sirois, and F. Grilli, "A new finite-element method simulation model for computing AC loss in roll assisted biaxially textured substrate YBCO tapes," *Superconductor Science and Technology,* vol. 23, no. 2, Art. no. 025001, 2009.

[27] L. N. Trefethen, *Approximation theory and approximation practice*. Philadelphia: SIAM, 2013.

[28] J. C. Mason and D. C. Handscomb, *Chebyshev polynomials*. Boca Raton: CRC press, 2002.

[29] A. D. Polyanin and A. V. Manzhirov, *Handbook of integral equations*. Chapman and Hall/CRC, 2008.

[30] I. S. Gradshteyn and I. M. Ryzhik, *Table of integrals, series, and products*. Academic press, 2014.

[31] M. Suenaga *et al.*, "Effects of a ferromagnetic substrate on hysteresis losses of a YBa2Cu3O7 coated conductor in perpendicular ac applied magnetic fields," *Physica C: Superconductivity,* vol. 468, no. 15, pp. 1714-1717, 2008.

[32] F. Rogier, "Mathematical and numerical study of a magnetostatic problem around a thin shield," *SIAM Journal on Numerical Analysis,* vol. 30, no. 2, pp. 454-477, 1993.

[33] J. W. Barrett and L. Prigozhin, "Bean's critical-state model as the p→∞ limit of an evolutionary p-Laplacian equation," *Nonlinear Analysis: Theory, Methods & Applications,* vol. 42, no. 6, pp. 977-993, 2000.

[34] C. Xue, "Private communication," ed, 2022.

[35] W. Norris, "Calculation of hysteresis losses in hard superconductors carrying ac: isolated conductors and edges of thin sheets," *Journal of Physics D: Applied Physics,* vol. 3, no. 4, p. 489-507, 1970.

[36] R. Duckworth, M. Gouge, J. Lue, C. Thieme, and D. Verebelyi, "Substrate and stabilization effects on the transport AC losses in YBCO coated conductors," *IEEE Transactions on Applied Superconductivity,* vol. 15, no. 2, pp. 1583-1586, 2005.